\documentclass[journal, twocolumn]{IEEEtran}
\usepackage{graphicx}
\usepackage{sidecap}
\usepackage{dcolumn}
\usepackage{bm}
\usepackage[utf8]{inputenc}
\usepackage[T1]{fontenc}
\usepackage{mathptmx}
\usepackage{textcomp}
\usepackage{siunitx}
\usepackage{bigints}
\usepackage{setspace}
\usepackage{hyperref}
\usepackage[switch]{lineno}
\usepackage[table,xcdraw,dvipsnames]{xcolor}
\usepackage{booktabs}
\usepackage{float}
\usepackage{color}
\usepackage{ulem}
\usepackage{multirow}

\begin{document}

\title{Magnetic anisotropy and GGG substrate stray field in YIG films down to millikelvin temperatures}

\author{\IEEEauthorblockN{Rostyslav~O.~Serha$^{1,2}$\IEEEauthorrefmark{1}, Andrey~A.~Voronov$^{1,2}$, David~Schmoll$^{1,2}$, Roman~Verba$^3$, Khrystyna~O.~Levchenko$^1$, Sabri~Koraltan$^{1,2,4}$, Kristýna~Davídková$^{1,2}$, Barbora~Budinska$^{1,2}$, Qi~Wang$^5$, Oleksandr~V.~Dobrovolskiy$^1$, Michal~Urbánek$^6$, Morris~Lindner$^7$, Timmy~Reimann$^7$, Carsten~Dubs$^7$, Carlos~Gonzalez-Ballestero$^8$, Claas~Abert$^{1,4}$, Dieter~Suess$^{1,4}$, Dmytro~A.~Bozhko$^9$, Sebastian~Knauer$^1$, and Andrii~V.~Chumak$^1$}
\\
\IEEEauthorblockA{
$^1$ Faculty of Physics, University of Vienna, 1090 Vienna, Austria.
\\
$^2$ Vienna Doctoral School in Physics, University of Vienna, 1090 Vienna, Austria.
\\
$^3$ Institute of Magnetism, Kyiv 03142, Ukraine.
\\
$^4$ Research Platform MMM Mathematics – Magnetism – Materials, University of Vienna, Vienna, Austria.
\\
$^5$ Huazhong University of Science and Technology, Wuhan, China.
\\
$^6$ CEITEC BUT, Brno University of Technology, 61200 Brno, Czech Republic.
\\
$^7$ INNOVENT e.V. Technologieentwicklung, 07745 Jena, Germany.
\\
$^8$ Institute for Theoretical Physics, Vienna University of Technology, 1040 Vienna, Austria.
\\
$^9$ Department of Physics and Energy Science, University of Colorado Colorado Springs, Colorado Springs, Colorado 80918, USA.
\\
Email: \IEEEauthorrefmark{1}rostyslav.serha@univie.ac.at}}
\maketitle
\begin{abstract}
Quantum magnonics investigates the quantum-mechanical properties of magnons such as quantum coherence or entanglement for solid-state quantum information technologies at the nanoscale. 
The most promising material for quantum magnonics is the ferrimagnetic yttrium iron garnet (YIG), which hosts magnons with the longest lifetimes. YIG films of the highest quality are grown on a paramagnetic gadolinium gallium garnet (GGG) substrate. 
The literature has reported that ferromagnetic resonance (FMR) frequencies of YIG/GGG decrease at temperatures below 50\,K despite the increase in YIG magnetization.
We investigated a 97\,nm-thick YIG film grown on 500\,µm-thick GGG substrate through a series of experiments conducted at temperatures as low as 30\,mK, and using both analytical and numerical methods.
Our findings suggest that the primary factor contributing to the FMR frequency shift is the stray magnetic field created by the partially magnetized GGG substrate. 
This stray field is antiparallel to the applied external field and is highly inhomogeneous, reaching up to 40\,mT in the center of the sample. 
At temperatures below 500\,mK, the GGG field exhibits a saturation that cannot be described by the standard Brillouin function for a paramagnet.
Including the calculated GGG field in the analysis of the FMR frequency versus temperature dependence allowed the determination of the cubic and uniaxial anisotropies. We find that the total anisotropy increases more than three times with the decrease in temperature down to 2\,K.
Our findings enable accurate predictions of the YIG/GGG magnetic systems behavior at low and ultra-low millikelvin temperatures, crucial for developing quantum magnonic devices.
\end{abstract}

\section{Introduction}
Magnonics is the field of science that deals with data carried and processed by spin waves and their quanta, magnons, in magnetically ordered media~\cite{Barman2021}. The ferrimagnet yttrium iron garnet (YIG) Y\textsubscript{3}Fe\textsubscript{5}O\textsubscript{12} is the material with the lowest known magnetic damping as bulk material~\cite{Dillon1957,LeCraw1958,Klingler2017} and in form of thin films~\cite{Dubs2017, Dubs2020, Ding2020, Heyroth2019, Cornelissen2015, Onbasli2014, Hahn2014}. Thus, YIG is the medium with the longest lifetimes of magnons up to one microsecond~\cite{SagaYIG}. Therefore, YIG has emerged as a preeminent material in RF technologies and magnonic experiments, showing promise for quantum magnonic applications. The field of quantum magnonics is a rapidly growing and highly promising research area that operates with quantum magnonic states, e.g. single magnons, and hybrid structures~\cite{Borst2023, Chumak2022, Awschalom2021, Lachance-Quirion2019, Lachance-Quirion2020, Xu2023, Li2020, Knauer2023, Yuan2022, Babak2022}. These investigations have to be performed at millikelvin temperatures to ensure that there is minimal thermal noise, allowing for precise observation and manipulation of the quantum states of magnons, which are extremely fragile to any kind of distortion.

Note, YIG was already the material of choice in experiments at low kelvin and ultralow millikelvin temperature magnonics for coupling to superconducting resonators~\cite{Borst2023, Novosad2022, Morris2017, Li2020} and propagating spin-wave spectroscopy~\cite{Karenovska2018, karenowska2015excitation, Knauer2023}. Furthermore, the first demonstration of magnon control and detection at the single magnon level~\cite{Lachance-Quirion2020} and the first measurements of the Wigner function of a single magnon~\cite{Xu2023} were performed also using a YIG sphere as a magnetic medium. Proposals for applications in quantum computing have also emerged, in particular the use of YIG spheres as magnonic transducers for qubits~\cite{Li2020}. A pressing question in the quantum magnonics community is how to bring the more complex, flexible, and property-rich structures employed in classical magnonic nanodevices to the quantum regime. Two dimensional geometries are a paramount example, for instance YIG films down to tens of nanometers thick grown on gadolinium gallium garnet (GGG) Gd\textsubscript{3}Ga\textsubscript{5}O\textsubscript{12} substrates~\cite{Dubs2020, Ding2020, Heyroth2019, Dubs2017, Onbasli2014, Hahn2014} enables the development of nanoscale magnetic devices and circuits~\cite{Chumak2022}. As the temperature decreases, the spin-wave damping in YIG/GGG increases up to tenfold due to various effects associated with impurities in YIG and parasitic influence of the paramagnetic GGG substrate~\cite{Jermain2017, Michalceanu2018, Kosen2019, Guo2022, Knauer2023, Cole2023}. Our experimental investigations of spin-wave damping agree well with the results reported in the literature (but are out of the scope of this article). 

Several experimental reports have demonstrated that lowering the temperature below about 50\,K in YIG/GGG shifts the ferromagnetic resonance (FMR) frequency or the frequency of propagating spin waves~\cite{Danilov1989, karenowska2015excitation, Wang2020B, Roos2022, Guo2022}. Our experimental results show the same behavior. The interpretation given by the authors in~\cite{Danilov1989}, agrees with our analysis. However, at the time of the study, the measurements and calculations were not performed below 4.2\,K, which is particularly interesting for quantum magnonics and is the temperature regime in which GGG shows a complex magnetic phase behavior~\cite{Petrenko1998, Deen2015}. Additionally, the role of the strong non-uniformity of the field induced by the partially magnetized GGG was not explored as well as the change of crystallographic anisotropy in YIG with decreasing temperature.

Here, through experiments, theory, and numerical simulations, we studied the stray field induced by the GGG substrate at temperatures as low as 30\,mK. Using vibrating-sample magnetometry (VSM), we measured the magnetization of GGG as a function of the applied field and temperature down to 2\,K. To extrapolate values for lower temperatures, we utilized the Brillouin function. We utilized analytical theory and numerical simulations to determine the GGG stray field and the cubic and uniaxial crystallographic anisotropies of YIG. Our findings revealed that the GGG field is strongly non-uniform, ranging from 10\,mT to 70\,mT for the YIG/GGG sample of (5x5)\,mm$^2$, 
 at a temperature of 2\,K, and an applied magnetic field of 600\,mT when the magnetization of GGG was 238\,kA/m. It has been experimentally confirmed by FMR measurements that the magnetization of the GGG substrate does not change with temperature below 500\,mK~\cite{Deen2015} in contrast to the Brillouin model of a paramagnet. 

\section{\label{Methods}Methodology}

\subsection{\label{ExpMethods}Experimental methods}

In our research, we studied a (5x5)\,mm$^2$ and 97\,nm-thick YIG film grown on a 500\,µm-thick GGG substrate, using liquid phase epitaxy~\cite{Dubs2017, Dubs2020}. We conducted stripline ferromagnetic-resonance (FMR) spectroscopy using a vector network analyzer (VNA), within a Physical Property Measurement System (PPMS), at temperatures ranging from 2\,K to 300\,K. FMR spectroscopy was conducted up to 40\,GHz. The experiments at ultralow temperatures were conducted in a dilution refrigerator that can reach a base temperatures of 10\,mK.

The following measurements were taken with the external magnetic field in the in-plane orientation and applied along the FMR stripline antenna. To determine the FMR spectrum for a specific field, $S_{12}$ and $S_{21}$ parameters were measured using a VNA not only at the target field but also at reference fields adjusted to approximately 15\,mT to 40\,mT, both above and below the desired value~\cite{Weiler2018}. By subtracting the averaged signals of the reference fields from the measured FMR signal, we obtained the FMR absorption spectrum in YIG that was not affected by GGG (see Fig.~\ref{f:2}~(b) as an example). This dual reference measurement approach enabled to obtain the best results when working with kelvin and sub-kelvin temperatures, since the GGG parasitic signal is greatly affected by the change in the applied field. To obtain the resonance frequency and full linewidth at half maximum, the background is first analyzed using a 1D cubic spline model~\cite{herrera2023double}. The resonance shape is then fitted using the double Lorentzian model, which individually describes the left and right sides of the asymmetric absorption peaks.

To accurately determine the magnetization of GGG for our analytic calculations and numerical simulations, we utilized vibrating-sample magnetometry (VSM) on a pure GGG slab in the temperature range from 2K to 300K. The Gd\textsuperscript{+3} ions in GGG have a relatively large spin (S\,=\,7/2), resulting in a saturation magnetization that is notably higher than that of YIG. Specifically, the saturation magnetization of GGG, denoted by $M_{\textup{GGG}}^\textup{s}$, is equal to 805 kA/m, as shown in Fig.~\ref{f:2}~(a).

Based on established practices, experimental FMR data can be used to extract the effective magnetization, the anisotropy fields, and the Gilbert damping parameter of a magnetic material (see~\cite{Boettcher2022} and the supplementary materials therein). In this study, the temperature-dependent saturation magnetization of YIG, $M_{\textup{YIG}}^{\textup{s}}$, is taken from the analytical calculation performed in~\cite{Hansen1974}. Thus, we can use this information to more accurately determine the anisotropy fields of YIG using them as a fitting parameter. 

As reported in the literature~\cite{Danilov1989, Knauer2023} and supported by our FMR analysis, the paramagnetic GGG becomes sufficiently magnetized at temperatures below about 100\,K, together with an external magnetic field applied. This magnetization induces a magnetic stray field $B_{\textup{GGG}}$ in the YIG layer, which causes a shift of the YIG FMR frequencies~\cite{Danilov1989}. For the in-plane applied magnetic field, the FMR shift is toward lower frequencies because $B_{\textup{GGG}}$ and the applied bias field $B_{\textup{0}}$ are antiparallel. Conversely, in an out-of-plane geometry, the stray field $B_{\textup{GGG}}$ aligns parallel to the field $B_{\textup{0}}$, resulting in a shift of the resonance frequency to higher values~\cite{Danilov1989}. The positive shift was also confirmed by our experimental results, but in this paper we focus only on the in-plane configuration. 

The magnitude of this inhomogeneous stray field $B_{\textup{GGG}}$ is influenced by both the temperature and the strength of the external magnetic field. At lower temperatures and higher external fields, the GGG-induced stray field becomes more pronounced, which is crucial for determining the FMR frequency $f_{\textup{FMR}}$. The modified Kittel formula tailored for in-plane magnetization geometry describes the influence of this field on the FMR frequency:
\begin{multline}\label{FMReq}
    f_{\textup{FMR}}=\gamma \cdot \sqrt{(B_{\textup{0}}-B_{\textup{GGG}})}\,\cdot\\\cdot\sqrt{(B_{\textup{0}}-B_{\textup{GGG}}+B_{\textup{ani}}^{\textup{eff}}+\mu_0\,M_{\textup{YIG}}^{\textup{s}})}\,\,,
\end{multline}

where $\gamma$ is the reduced gyromagnetic ratio, $B_{\textup{0}}$ is the applied external field, $B_{\textup{ani}}^{\,\textup{eff}}$ the effective crystallographic anisotropy field \cite{Boettcher2022}, which is a fit parameter as described in detail below, $B_{\textup{GGG}}$ is the GGG-induced stray field defined analytically in the next section and $M_{\textup{YIG}}^{\textup{s}}$ the theoretical value for the saturation magnetization of YIG at any temperature taken from~\cite{Hansen1974}.

\subsection{\label{AnaMethods}Analytical calculation of GGG field}

At low temperatures below 10\,K and in the presence of magnetic fields measuring several hundred millitesla, GGG attains a significant magnetization  exceeding hundreds of kA/m (see Fig.\,\ref{f:2}\,(a)). In this state, the GGG substrate becomes a magnet and emits a stray magnetic field that expands beyond its volume. To accurately determine the strength and characteristics of the stray field generated by the GGG, two crucial parameters are required: $M_{\textup{GGG}}$, which denotes the magnetization of GGG, and $\tilde N_{xx}$, which represents the averaged mutual in-plane demagnetization factor of GGG and YIG slabs. Then, one can accurately determine the strength and characteristics of the stray field generated by the GGG in the direction of the external field $B_{\textup{GGG}}$ as
\begin{equation}\label{BGGG}
    B_{\textup{GGG}}=\mu_{0}\,M_{\textup{GGG}}(T,B_{\textup{0}})\,\tilde N_{xx}\,\,.
\end{equation}
$B_{\textup{GGG}}$ is crucial for understanding its overall magnetic influence, particularly at low temperatures, on the internal field of the YIG film. Net GGG magnetization $M_{\textup{GGG}}$ is given by the implicit equation~\cite{Barak1992}:
\begin{multline}\label{MGGG}
    M_{\textup{GGG}}=M^{\textup{s}}_{\textup{GGG}}\cdot\\\cdot\mathcal{B}_{\frac{7}{2}}\biggl(\frac{7}{2}\cdot\frac{g\,\mu_{\textup{B}}\cdot(B_0-\mu_0\,M_{\textup{GGG}}\,N_{xx}+\lambda\,\mu_0\,M_{\textup{GGG}})}{k_{\textup{B}}\,T}\biggr)\,\,,
\end{multline}
where $M^{\textup{s}}_{\textup{GGG}}$\,=\,805\,kA/m denotes the saturation magnetization of GGG, $g$\,=\,2 the Landé factor, $\mu_{\textup{B}}$ the Bohr magneton, $\mathcal{B}_{\frac{7}{2}}$ the Brillouin function for the angular momentum $J\,$=$\,\frac{7}{2}$, $\lambda$ the coefficient of the molecular field and $N_{xx}$ is the standard demagnetization factor (i.e. self-demagnetization) of GGG sample. 

In our case of thin GGG cuboid of the sizes $2a \times 2a \times 2c$, where $c \ll a$, self demagnetization factor can be approximated as \cite{Aharoni1998}
\begin{equation}
    N_{xx}\approx \frac{c}{\pi a}\left(0.726-\log \frac{c}{a} \right) \ .
\end{equation}
For the mutual demagnetization factor, we account that YIG film is much thinner than GGG, and that we deal with nonuniform FMR in YIG, because the lateral sizes of YIG are much larger than spin-wave free path, so that standing eigenmodes are not formed. In this case, the FMR peak position is mostly determined by central YIG area, for which mutual demagnetization factor is  
\begin{equation}
    \tilde N_{xx}= \frac{1}{\pi}\cdot \arctan\left[\frac{\sqrt{2} c}{\sqrt{a^2 + 2 c^2}}\right]\ .
\end{equation}
Full expressions for arbitrary prisms are available in~\cite{Aharoni1998}. Note, both approximations for self and mutual demagnetization factors may not be adequate for small external fields $B_0$, as it neglects the potential nonuniform magnetization of GGG.

To ensure that our calculations and numerical micromagnetic simulations would be most precise, the magnetization of GGG $M_{\textup{GGG}}$ was measured for each temperature within the same field range used in the FMR experiments. For calculations at temperatures below 2\,K, we relied on the VSM data obtained earlier. The data was fitted using Eq.\,\ref{MGGG} by adjusting the molecular field $\lambda$\,=\,-0.854 and saturation magnetization $M^{\textup{s}}_{\textup{GGG}}$\,=\,756\,kA/m as fitting parameters. This fitting method was used to extrapolate the values of $M_{\textup{GGG}}$ with respect to temperature $T$ and external magnetic field $B_0$. It allowed us to compare our theoretical and simulation work with experimental results even at temperatures below 2\,K.

\subsection{\label{NumMethods}Numerical simulations}

To gain a better understanding of the stray magnetic field present in the YIG layer on the GGG substrate, we also conduct micromagnetic simulations using magnum.np~\cite{bruckner2023magnum} and mumax$^3$~\cite{vansteenkiste2014design}. These simulations offer a more precise depiction of the inhomogeneity of the field.

The simulations for the GGG substrate were conducted using the same geometric parameters as in the experimental study. The saturation magnetization, denoted as $M_{\mathrm{GGG}}$, was take from the experimental results measured using VSM. Although micromagnetic solvers are not intended for simulations of a paramagnetic and cannot give a rigorous description (it requires other numerical approaches \cite{Smith2010}, which are too cumbersome in 3D), static paramagentic behavior of the material can be reasonably simulated by setting the exchange constant $A_\mathrm{ex}$ to zero. Due to the small magnetic susceptibility $\chi$ (see Ref.\,\cite{Deen2015}) in the linear regime, it is assumed that the magnitude of the local magnetization does not change by cause of self-demagnetizing effects. The external magnetic field $B_0$ was applied in-plane, as depicted in Fig.\,\ref{f:1}\,(a). To ensure saturation within a simulation time of 2\,ns, a Gilbert damping factor of $\alpha$\,=\,$1$ was utilized, and the resulting demagnetization field was recorded as a projection in the direction of the external bias field $B_\mathrm{GGG}$. The simulation employed a mesh of $1000 \times 1000 \times 26$ cells. Given that the simulation's objective was to examine the magnetization's relaxed state in the absence of exchange interaction, a cell dimension of ($5 \times 5 \times 20$)\,µm$^3$ was deemed adequately sufficient. In the end, to determine the strength of the demagnetization field $B_\mathrm{GGG}$ at the interface between YIG and GGG, the mean-field value from the last two layers along the $z$-axis was calculated.
    
\section{Results}
\begin{figure}[ht!]
    \includegraphics[width=1\linewidth]{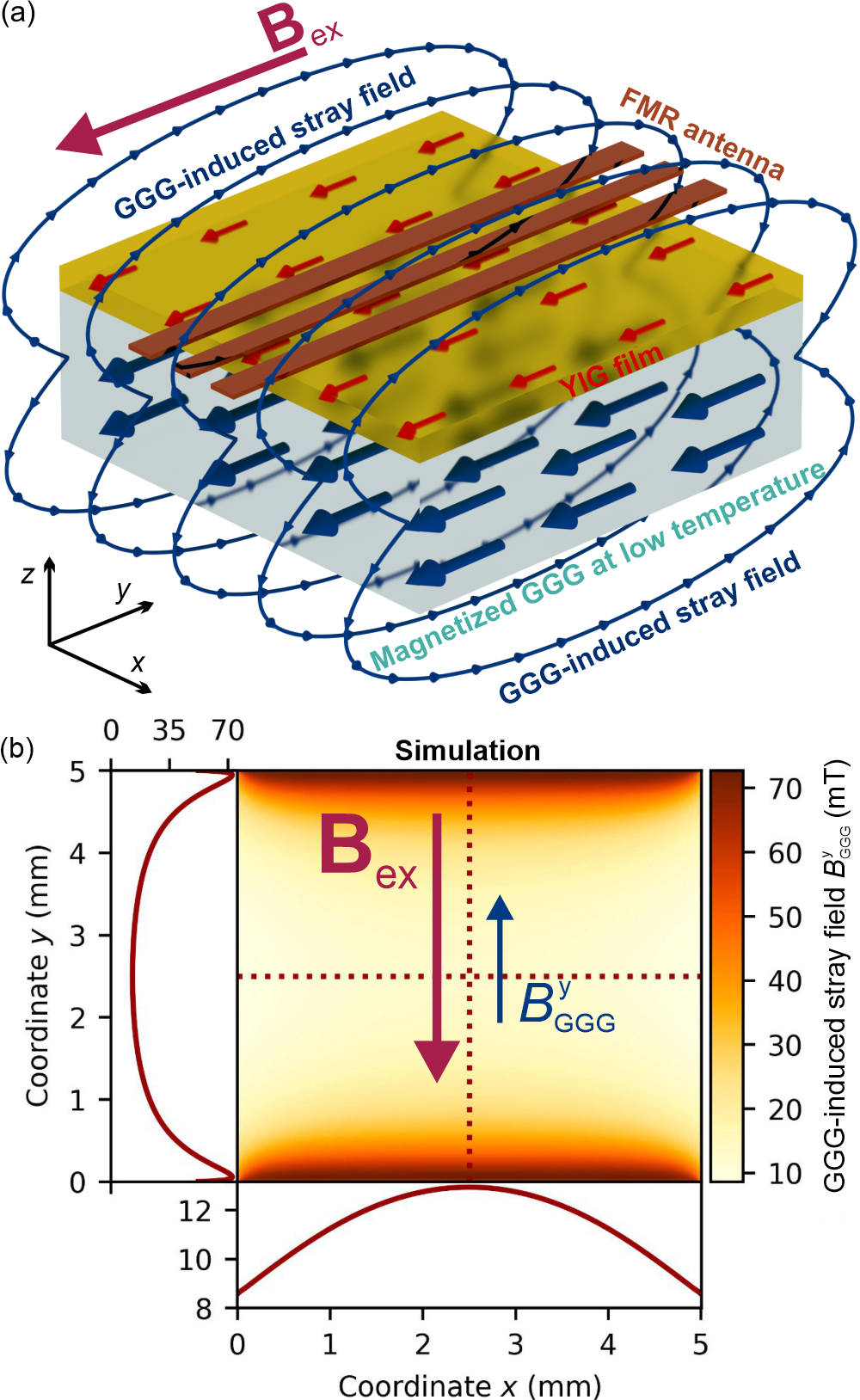}
    \caption{(a) Depiction of the experimental system of a YIG film grown on GGG. The sample is in-plane magnetized by an external magnetic field, and at temperatures approaching 0\,K the paramagnetic GGG spin system saturates. The substrate creates an inhomogeneous stray field, that becomes an additional component to the internal magnetic field of the YIG but is oriented antiparallel to the external field. For measuring FMR the YIG film sample is placed on a CPW antenna, through which the magnetic system is excited. (b) Micromagnetic simulation of the highly inhomogeneous $B_{\textup{GGG}}^\textup{y}$ stray field y-component at the interface between YIG and GGG layers. The inhomogeneity can also be seen for the $x$ and $y$-axis for the center of the sample by the two cross-section plots marked in the 2D map with the red dotted lines. The simulation is performed using mumax$^3$~\cite{vansteenkiste2014design} for the temperature 2\,K and the strength of the applied external magnetic field of 600\,mT at which the magnetization of GGG was 238\,kA/m.}
    \label{f:1}
    		
\end{figure}

\begin{figure}[p]
    \includegraphics[width=1\linewidth]{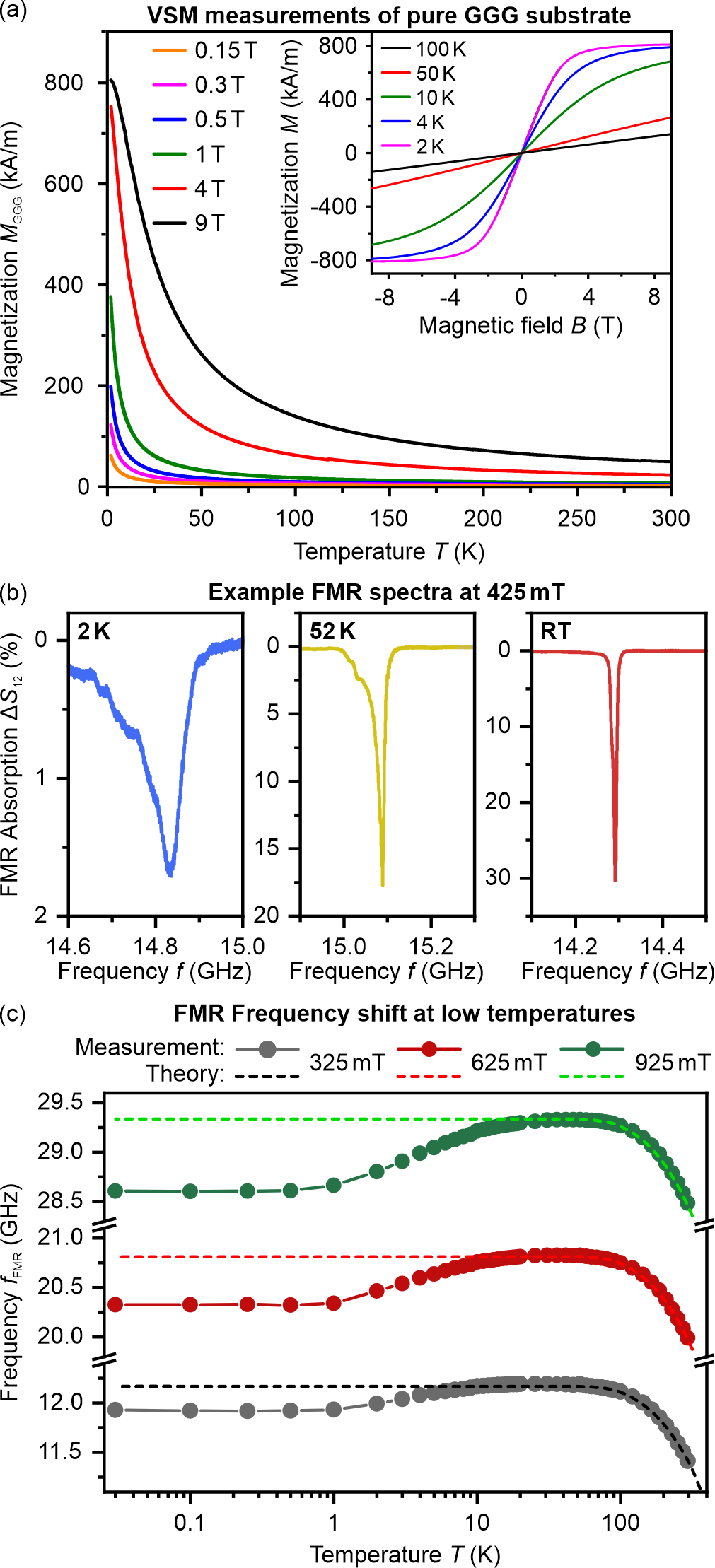}
    \caption{(a) VSM measurements of the magnetization of GGG $M_{\textup{GGG}}$ as function of the temperature $T$ and the external magnetic field $B_0$. $M_{\textup{GGG}}$ saturates at a value of 805\,kA/m. The graphs represent the GGG magnetization over the range available in the experiment. (b) Example spectra of FMR at the temperatures of 2\,K, 52\,K, and room temperature (RT) for an external magnetic field of 425\,mT. (c) FMR frequency as function over the temperature, plotted in a $x$-axis logarithmic scale, for 3 different magnetic fields. Measurements are depicted as points and were performed with the sample magnetized in the $\langle1\overline{1}0\rangle$ direction. The dotted lines are portraying the analytical calculation for the frequency of the FMR by the Kittel formula, which is neglecting the induced GGG stray field. The parameters for the gyromagnetic ratio $\gamma$ and effective anisotropy field $B_{\textup{ani}}^{\,\textup{eff}}$ are obtained by fitting room temperature measurements.}\label{f:2}
\end{figure}

To simulate the strength and profile of the stray field induced by the paramagnetic substrate over the YIG layer, we first measured the magnetization of a bare GGG substrate, as shown in Fig.\,\ref{f:2}\,(a). In the partially magnetized state, GGG has a small value of magnetic susceptibility $\chi\,\approx$\,0.3 and creates a highly inhomogeneous y-field component $B_{\textup{GGG}}^\textup{y}$ on its surface, as seen in Fig.\,\ref{f:1}\,(b). As an example, at 2\,K, this induced field $B_{\textup{GGG}}^\textup{y}$ opposes the external magnetic field of 600\,mT. It varies in strength from 13\,mT at the center to 70\,mT at the edges, antiparallel to the direction of the external magnetic field. This field is crucial in investigating YIG/GGG systems at low temperatures.

Fig.\,\ref{f:2}\,(b) displays two spectra that exemplify the impact of cryogenic temperatures on the FMR signal of the YIG film in an external magnetic field of 425\,mT. The FMR peak is almost entirely Lorentzian-shaped at room temperature. However, when the temperature decreases below 100\,K, the peak becomes asymmetric on one side, as shown in Fig.~\ref{f:2}~(b) for 52\,K, and then broadens significantly with a lower resonance frequency, as demonstrated for 2\,K. Additionally, the amplitude of the FMR peak decreases as the temperature reduces. Note the shape and width of the FMR peak will be analyzed quantitatively and discussed in future work. For now, this manuscript exclusively focuses on the internal field and the FMR frequency of the YIG film.
 
Fig.\,\ref{f:2}\,(c) clearly shows the impact of the stray field induced by the GGG substrate. The graph displays the FMR frequencies $f_{\mathrm{FMR}}$ obtained for the YIG film at various temperatures, ranging from 300\,K to 30\,mK. As the temperature decreases, the FMR frequency increases due to the rise in the saturation magnetization of YIG. The theoretical curves, represented by dashed lines, were calculated using Eq.\,\ref{FMReq}, with the gyromagnetic ratio $\gamma$ and effective anisotropy field $B_{\textup{ani}}^{\,\textup{eff}}$ taken from the Kittel fit at room temperature, but neglecting the contribution of the GGG-induced stray field ($B_{\textup{GGG}}$\,=\,0).

At temperatures below 50\,K, the experimental FMR frequencies deviate from the theoretical values due to the GGG-induced stray field. The data indicate that for an external field $B_0$ of 925\,mT, the FMR frequency begins to decrease at approximately 50\,K, while for $B_0$ values of 325\,mT, this occurs at around 25\,K. At temperatures below 2\,K and high external fields of 925\,mT, there is a notable deviation of over 0.5\,GHz between the experimental and theoretical results. This difference is attributed to the GGG stray field, which opposes the external magnetic field and lowers the FMR frequency $f_{\mathrm{FMR}}$ of YIG. The FMR frequency shift, dependent on $M_{\textup{GGG}}$, becomes more pronounced as the temperature decreases and the external magnetic field increases, as shown in Fig.\,\ref{f:2}\,(c). The results of comparing three different magnetic fields at 30\,mK shows that the impact is more pronounced at higher excitation frequencies, which is associated with stronger magnetic fields.

\begin{figure}[p]
    \includegraphics[width=1\linewidth]{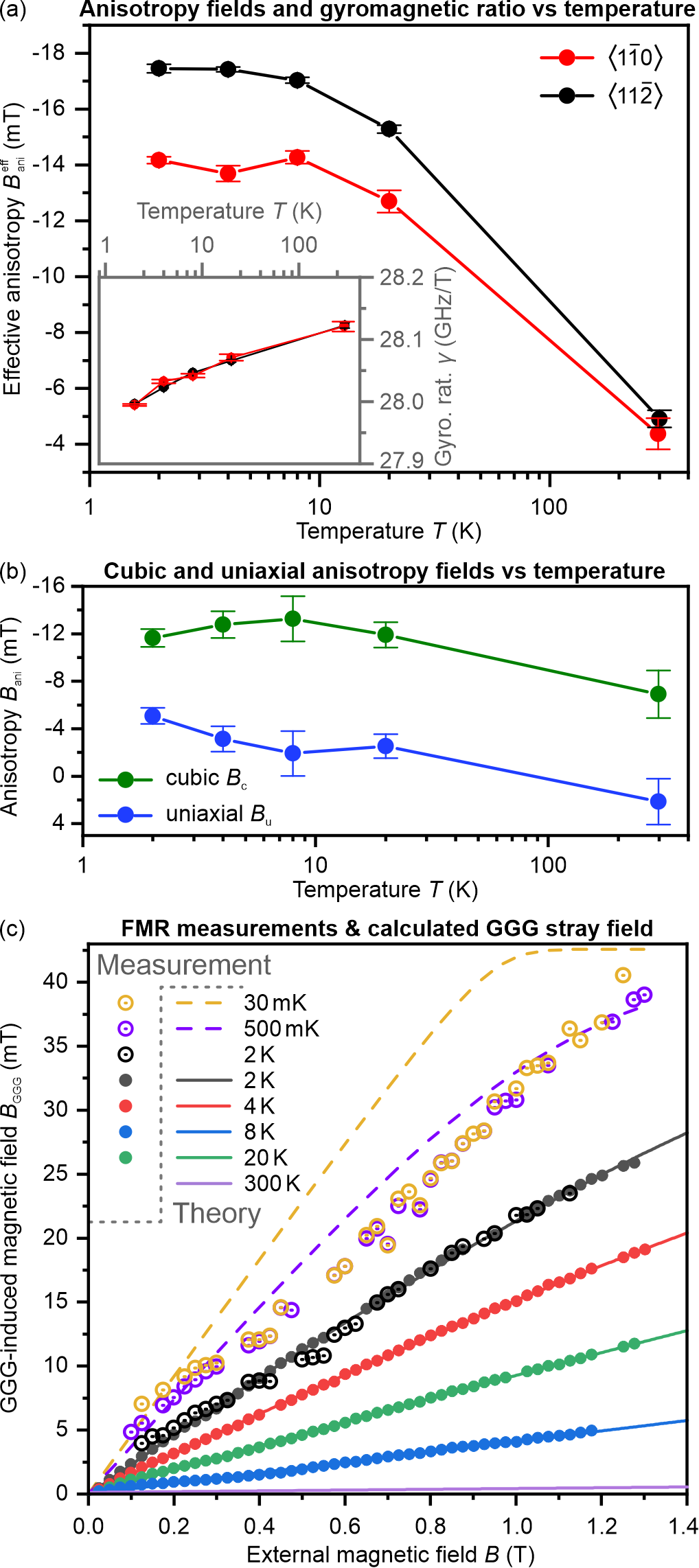}
    \caption{(a) Effective anisotropy field $B_{\textup{ani}}^{\,\textup{eff}}$ of the YIG film as a function of the temperature in a $x$-axis logarithmic scale for two different crystallographic magnetization directions --- $\langle1\overline{1}0\rangle$ and $\langle11\overline{2}\rangle$. The inset depicts the gyromagnetic ratio $\gamma$ as a function of the temperature for the same magnetization geometries respectively. The points are obtained as fitting parameters from Eq.\,\ref{FMReq}. (b) Cubic and uniaxial anisotropiy fields $B_{\textup{c}}$, $B_{\textup{u}}$ vs temperature $T$. Values for $B_{\textup{c}}$ and $B_{\textup{u}}$ were obtained from according Eq.\,\ref{FMR110} and \ref{FMR112} for the Kittel fits~\cite{Bobkov1997}. (c) GGG-induced stray field as function of the externally applied magnetic field $B_0$. Solid lines are the values calculated from the measured $M_{\textup{GGG}}$ by~Eq.\,\ref{BGGG} of the VSM and solid lines are values calculated from the FMR peaks measured in the PPMS by converting Eq.\,\ref{FMReq} for $B_{\textup{GGG}}$. The hollow points are $B_{\textup{GGG}}$ values calculated from FMR peak measurements performed in the dilution refrigerator. The dotted lines are calculated via~Eq.\,\ref{BGGG} by taking the extrapolated values of $M_{\textup{GGG}}$ by the Fit from Eq.\,\ref{MGGG}.}
    \label{f:3}
\end{figure}

Below 500\,mK, the frequency $f_{\mathrm{FMR}}$ displays unusual behavior as its decline stabilizes, showing negligible change down to 30\,mK, despite the varied magnetization of GGG in this temperature range as predicted by the Brillouin function (Eq.,\ref{MGGG}). This phenomenon is due to the complex nature of GGG, which possesses a geometrically highly frustrated spin system \cite{Schiffer1994, Tsui1999}, leading to a complex phase diagram for temperatures below 1\,K \cite{Petrenko1998, Deen2015}. Understanding the sub-kelvin temperature behavior of GGG requires considering competing interactions among loops of spins, trimers, and decagons, along with the interplay between antiferromagnetic, incommensurate, and ferromagnetic orders \cite{Deen2015}. Consequently, the Brillouin function fails to describe the magnetization of GGG below 500\,mK and can be seen even clearer in later described Fig.\,\ref{f:3}\,(c). This finding aligns with previous experimental studies \cite{Deen2015} using different techniques, such as single-crystal magnetometry and polarized neutron diffraction.
Incorporating the analytically-calculated GGG stray field $B_{\textup{GGG}}$ into the Kittel formula Eq.\,\ref{FMReq} is necessary for determining the FMR frequency. To accurately identify the FMR frequencies $f_{\mathrm{FMR}}$ e.g. shown in  Fig.\,\ref{f:2}\,(b), it is essential to account for the GGG-induced stray field $B_{\textup{GGG}}$ at the sample center. At this location, the gradient is zero, and the region of the same field magnitude is the largest, most significantly affecting the area excited by the microwave stripline and determining the position of the FMR peak. The red vertical dashed line in Fig.\,\ref{f:1}\,(b) approximately depicts the position of the microstrip FMR antenna. Once incorporated into the Kittel equation, the gyromagnetic ratio $\gamma$ and the effective anisotropy field $B_{\textup{ani}}^{\,\textup{eff}}$ become fitting parameters.
Obtaining $B_{\textup{ani}}^{\,\textup{eff}}$ allows us to gain insight into and make predictions about the internal field of thin YIG films at low temperatures. Fig.\,\ref{f:3}\,(a) displays the fitting outcomes for the magnetization orientations of $\langle1\overline{1}0\rangle$ (red) and $\langle11\overline{2}\rangle$ (black), with the effective anisotropy field $B_{\textup{ani}}^{\,\textup{eff}}$ plotted against temperature on a $x$-axis logarithmic scale. The error bars are taken from the root-mean-square deviation of the fit. At room temperature, the anisotropy field is relatively small, measuring approximately 5\,mT, with a variation of about 0.6\,mT between the two orientations due to cubic anisotropy of the YIG single crystal.

However, as the temperature decreases, both the strength of $B_{\textup{ani}}^{\,\textup{eff}}$ and the difference between $\langle1\overline{1}0\rangle$ and $\langle11\overline{2}\rangle$ magnetization directions increase significantly. At a temperature of 2\,K, the effective anisotropy field $B_{\textup{ani}}^{\,\textup{eff}}$ is more than 3.2 and 3.5 times larger for the $\langle1\overline{1}0\rangle$ and $\langle11\overline{2}\rangle$ directions, respectively, than at room temperature. The inset in Fig.\,\ref{f:3}\,(a) displays the second fitting parameter, the gyromagnetic ratio $\gamma$, as a function of temperature on a $x$-axis logarithmic scale. It is known from previous research on YIG that  $\gamma$ is considered to be weakly temperature  dependent~\cite{Haidar2015,Maier2017s,Cole2023}. The behavior of $\gamma$ shows a very weak decrease at lower temperatures, changing from 28.13\,GHz/T to 28.0\,GHz/T. $\gamma$ is nearly identical for both magnetization directions, and the values fall within each other's error bars.

To get a better understanding of the changes in the effective anisotropy field $B_{\textup{ani}}^{\,\textup{eff}}$, the field was divided into its two main components: the cubic anisotropy field $B_\textup{c}$ and the uniaxial anisotropy field $B_\textup{u}$. This separation is achieved by considering the two magnetization directions and applying two separate equations for the FMR frequency, as described in~\cite{Bobkov1997}.
\begin{multline}\label{FMR110}
f_{\textup{FMR}}^{\langle1\overline{1}0\rangle}=\gamma^{\langle1\overline{1}0\rangle} \cdot \sqrt{(B_{\textup{0}}-B_{\textup{GGG}})}\,\cdot\\\\
\cdot\sqrt{(B_{\textup{0}}-B_{\textup{GGG}}-B_{\textup{u}}-B_{\textup{c}}+\mu_0\,M_{\textup{s}})-2B^2_{\textup{c}}}\,\,,
\end{multline}
\begin{multline}\label{FMR112}
f_{\textup{FMR}}^{\langle11\overline{2}\rangle}=\gamma^{\langle11\overline{2}\rangle} \cdot \sqrt{(B_{\textup{0}}-B_{\textup{GGG}})}\,\cdot\\\\
\cdot\sqrt{(B_{\textup{0}}-B_{\textup{GGG}}-B_{\textup{u}}-B_{\textup{c}}+\mu_0\,M_{\textup{s}})}\,\,.
\end{multline}
Using Eq.\,\ref{FMR110} and Eq.\,\ref{FMR112} to fit the FMR frequency $f_{\mathrm{FMR}}$ data, and utilizing the gamma values from Fig.\,\ref{f:3}\,(a), we can obtain values for the two distinct anisotropy fields, $B_\textup{c}$ (green) and $B_\textup{u}$ (blue) and their errors by root-mean-square deviation. The resulting plot in Fig.\,\ref{f:3}\,(b) shows the anisotropy field $B_\textup{ani}$ as a function of temperature $T$ on a $x$-axis logarithmic scale. Our experimental findings at room temperature indicate (-6.9\,\textpm\,2)\,mT for $B_\textup{c}$ and (2.1\,\textpm\,2)\,mT for $B_\textup{u}$, which agree well with previously reported values for thin YIG films~\cite{Dubs2020}. Notably, the cubic anisotropy increases to (-11.6\,\textpm\,0.8)\,mT at temperatures as low as 2\,K, almost doubling the room temperature value. The uniaxial anisotropy exhibits a unique characteristic of changing its positive-to-negative sign at cryogenic temperatures and reaching a peak value of (-5\,\textpm\,0.7)\,mT at 2\,K. 

With this understanding of the low-temperature anisotropy and the stray field caused by GGG, we can make accurate predictions about the FMR behavior in YIG films even at temperatures as low as 2\,K. Note, that the anisotropy increase reaches a saturation point below 10\,K, as demonstrated in Fig.\,\ref{f:3}\,(a). As a result, we can assume that the anisotropy remains constant down to the millikelvin temperatures and can be treated as equivalent to the 2\,K values. 

However, in the absence of magnetization $M_{\textup{GGG}}$ measurements below 2\,K, we cannot utilize experimental data to compute the stray field $B_{\textup{GGG}}$ using Eq.\,\ref{BGGG}. Therefore, we must rely on the Brillouin fit presented in Sec.\,\ref{ExpMethods} (Eq.\,\ref{MGGG}) and extrapolate $M_{\textup{GGG}}$ as an estimation. Additionally, a second method was used to determine the values for $B_{\textup{GGG}}$ at millikelvin temperatures in the center of the YIG sample by measuring the FMR position at these temperatures and rearranging Eq.\,\ref{FMReq} to solve for $B_{\textup{GGG}}$.
The results are shown in Fig.\,\ref{f:3}\,(c). It depicts the stray field induced by GGG at the center of the sample as a function of the externally applied magnetic field $B_0$. The solid lines represent the calculated values of $B_{\textup{GGG}}$ for temperatures of 2\,K and above, while the solid points represent the obtained $B_{\textup{GGG}}$ from the FMR measurements. Both data sets match perfectly.
This alignment, in Eq.\,\ref{FMReq}, highlights the accuracy of the fitted effective anisotropy $B_{\textup{ani}}^{\,\textup{eff}}$, affirming the precision and appropriateness of the fit. The data points shown by the hollow-center-dot illustrate the $B_{\textup{GGG}}$ values acquired via FMR measurements in the dilution refrigerator, spanning a temperature range of 30\,mK to 2\,K.
It is worth noting that the values of $B_{\textup{GGG}}$ at 2\,K obtained from both the dilution refrigerator and the PPMS measurements are consistent. 

The extrapolated data obtained through the Brillouin fit is shown as dotted lines for both 500\,mK and 30\,mK, in comparison to the experimental data (Fig.,\ref{f:3}\,(c)). At 500\,mK, the extrapolation matches at fields above 1\,T and below 300\,mT with the experimental data but diverts in-between (purple circle points and dashed line). At 30\,mK, it does not match (yellow circle points and dashed line).
The extrapolated field strength of GGG (dashed yellow) experiences a sharp increase and then saturates above 1.1\,T. 
While the extrapolated curves deviate, the measured data for 30\,mK and 500\,mK overlap.
These results show that the method of fitting the GGG magnetization $M_{\textup{GGG}}$ and the stray magnetic field $B_{\textup{GGG}}$ effectively describes the inter- and extrapolations only at temperatures above 500\,mK. 
This limitation confirms the conclusion above. $M_{\textup{GGG}}$ is solely dependent on the externally applied magnetic field below 500\,mK, which is supported by the behavior of the FMR frequency in Fig.\,\ref{f:2}\,(c) and previous research on the complex behavior of GGG phase states~\cite{Deen2015}. At these temperatures, GGG was observed to transition through various magnetic phases, such as spin glass and antiferromagnetic, depending on the external magnetic field.~\cite{Petrenko1998}. 

\section{Conclusion}
We found that YIG films grown on GGG substrates are impacted by stray fields originating from the partially magnetized paramagnetic GGG at low temperatures and under externally applied magnetic fields. The strength and configuration of these fields depend on the shape of the GGG substrate (the ratio between the width, length and thickness of the sample) and are highly inhomogeneous across the YIG layer. In the in-plane magnetization geometry, the stray field can reach up to 40\,mT in the center and increases five-fold at the edges of the sample. 

We used an analytical approach validated by micromagnetic simulations to calculate the stray field $B_{\textup{GGG}}$ induced by GGG. This approach allowed us to integrate $B_{\textup{GGG}}$ into the Kittel-fit formula and accurately determine the effective anisotropy field $B_{\textup{ani}}^{\,\textup{eff}}$ in the crystallographic directions $\langle1\overline{1}0\rangle$ and $\langle11\overline{2}\rangle$ of the YIG film for temperatures as low as 2\,K. Moreover, we were able to extract the crystallographic cubic and uniaxial anisotropy fields, $B_{\textup{c}}$ and $B_{\textup{u}}$, respectively. These fields increase in magnitude from -6.9\,mT and 2.1\,mT at room temperature to -11.6\,mT and -5\,mT at 2\,K.
The anomalous behavior of the FMR frequency of YIG, which is constant for temperatures below 500\,mK, can be explained by the absence of the variation of the GGG magnetization $M_{\textup{GGG}}$ with decreasing temperature, and therefore by the GGG-induced magnetic field $B_{\textup{GGG}}$. This behavior can be described by the property of GGG as a geometrically highly frustrated magnet, resulting in the complex phase transition diagram of GGG at these temperatures and fields~\cite{Schiffer1994, Tsui1999, Petrenko1998, Deen2015}. Our findings enable accurate predictions of the YIG/GGG magnetic behavior at low and ultra-low temperatures, which is a key element for successfully implementing YIG/GGG quantum-magnonic networks.

\section*{acknowledgements}
AVC acknowledges the Austrian Science Fund FWF for the support by the project I-6568 "Paramagnonics".
SK acknowledges the support by the H2020-MSCA-IF under Grant No. 101025758 ("OMNI").
DS acknowledges the Austrian Science Fund FWF for the support by the project project I 4917-N "MagFunc".
KOL acknowledges the Austrian Science Fund FWF for the support through ESPRIT Fellowship Grant ESP 526-N "TopMag".
Work by DAB was supported by the U.S. Department of Energy (DOE), Office of Science, Basic Energy Sciences (BES) under Award DE-SC0024400.
RV acknowledges support of the NAS of Ukarine (project 0123U104827).
The work of ML was supported by the German Bundesministerium
für Wirtschaft und Energie (BMWi) under Grant
No. 49MF180119. CD thanks R. Meyer (INNOVENT e.V.) for technical support.
CGB acknowledges the Austrian Science Fund FWF for the support with the project PAT-1177623 "Nanophotonics-inspired quantum magnonics". The authors thank Prof. Dr. G. A. Melkov for valuable scientific discussions.

\section*{Author contributions}
ROS conducted all measurements, processed and analyzed the data, and authored the initial draft of the manuscript.
AAV executed the micromagnetic simulations, implemented the method for fitting the data, and contributed to data analysis. SAK, CA and DS developed the software used for micromagnetic simulations.
DAS, SAK, DAB and SK constructed the experimental setup and supported the experimental investigations.
KOL, KD, QW and MU assisted in interpreting the experimental data and provided insights into the measurement results.
ML, TR and CD synthesized the YIG film. 
RV and CGB provided theoretical support.
BB and OVD facilitated support in conducting PPMS measurements. SK supervised the millikelvin experiments. AVC led the project. All authors discussed results and contributed to the manuscript. 

\section*{Competing interests}
The authors declare no competing interests.


\textbf{Data availability}
The data that support the findings of this study are available
from the corresponding author upon reasonable request.
\emergencystretch=5em
\bibliographystyle{naturemag}
\bibliography{BIB}


\end{document}